# Robustness of Link-prediction Algorithm Based on Similarity and Application to Biological Networks

Liang Wang, Ke Hu and Yi Tang[1]

*Department of Physics and key laboratory of Intelligent Computing & Information Processing (Xiangtan University) Ministry of Education, Xiangtan University, Xiangtan 411105, Hunan, China*

**Abstract**: Many algorithms have been proposed to predict missing links in a variety of real networks. These studies focus on mainly both accuracy and efficiency of these algorithms. However, little attention is paid to their robustness against either noise or irrationality of a link existing in almost all of real networks. In this paper, we investigate the robustness of several typical node-similarity-based algorithms and find that these algorithms are sensitive to the strength of noise. Moreover, we find that it also depends on network's structure properties, especially on network efficiency, clustering coefficient and average degree. In addition, we make an attempt to enhance the robustness by using link weighting method to transform un-weighted network to weighted one and then make use of weights of links to characterize their reliability. The result shows that proper link weighting scheme can enhance both robustness and accuracy of these algorithms significantly in biological networks while it brings little computational effort.

## 1. INTRODUCTION

Many social, biological and communication systems can be properly described by complex networks whose nodes represent individuals or organizations and links mimic the interactions among them. Thus, complex networks have attracted a great deal of attention and become a powerful tool to analyze many different kinds of complex system. Up to now, the study of complex networks has gained vast progress in understanding the structure, evolution and function of these systems [1-5]. Recently, another important issue relevant to complex networks is the link prediction which involves estimating the likelihood of the existing yet unknown connections between individuals. Such kind of problems has caught increasing attention due to both theoretical significance and potential application [6-9]. On the one hand, link prediction can provide significant instruction for mining unknown connections in incomplete networks. For example, in protein-protein interaction network, a certain number of interactions are not tested, which may be very important. In order to find the missing links, especially when the network is huge and sparse, checking every potential interaction blindly in the laboratory is time consuming as the number of all possible links increases squarely with that of nodes [6]. And large numbers of samples often mean high cost which is usually impossible. If we can detect the underlying links which are most likely to exist by adopting some link prediction methods, the experiments will be well targeted to identify the missing links. On the other hand, it can help us to predict the links that may exist in the future along with evolution of networks. For example, in on-line social networks [10, 11], very likely but not yet existent links can be recommended to become promising friendships, which can help users to find new friends and enhance our loyalties to the networks. In addition, study of link prediction is helpful to fulfill some other tasks, such as understanding the evolution mechanisms of networks and the analysis of the networked structures [12, 13].

---

[1] Corresponding author. Email address: tangyi@xtu.edu.cn

In the past few years, a large number of algorithms have been developed for the link prediction in complex networks. Zhou *et al.* systematically summarized recent progress about link prediction algorithms [14]. Traditional units-attribute-based link prediction has been applied to several special networked systems, but success has been limited due to unavailability of content and attributes information. Additionally, large numbers of units and inter actions with distinct natures among various large-scale complex systems often need to analyzed. This labor-intensive task results in the lack of a general and effective approach. Fortunately, rapid advance in network theory provides us with new avenues for developing effective and efficient algorithms to predict missing links directly from the network topologies. In this stage, many prediction algorithms have been recently proposed and their successes have confirmed that a serious consideration of topological characteristics of network may indeed provide useful information and insights for link prediction.

In generally, the link prediction algorithms based on network topologies are designed according to the measures of the structural similarity of nodes, which can be classified as local and global methods. Because of only considering local information, the computational efforts of local methods is far less than global methods, especially in these networks with large scale and sparse topological structure, this advantage is more obvious. Among a number of indices based on local information, Liben-Nowell and Kleinberg [15] showed that the Common Neighbors (CN) [16] and Adamic-Adar (AA) [17] index are the best. Zhou *et al.* [18] compared a number of local similarity index on some real networks, and proposed two new indexes named Resource Allocation (RA) and the Local path (LP) index. The study found that the two new indexes are significantly better in the ability of prediction than the other known nine kinds of index based on local information similarity. Lü *et al.* [19] further analyzed the performance of local path index in the network model with controlled noise intensity and network density, and found that the prediction ability of LP index can rival global methods, such as the Katz index [20], when the network average distance is small, and even under the condition of large noise, the LP index has higher precision than the Katz index.

Both local and global methods depend on topological information of network, but these informations at many times are not accurate. In protein-protein interaction networks, with the development of high throughput testing technology, the data of large-scale protein interaction is keeping accumulated. But the quality of these data is severely affected for a large number of false positive and negative noise, a systematic comparison of several high-throughput method of reference high-quality data set showed that these methods have accuracies below 20% [21]. In addition, missing data due to individual no response and dropout [12], informant inaccuracy [22] and sampling biases [13] in social networks may mislead us to get inaccuracies topological properties, while they are pervasive. Additionally, we will get different topological properties for a network with different methods. Therefore we have to consider the noise influence to the precision of algorithm, while little attention has been paid on this by now.

In this paper, we study the robustness of algorithm in some real networks. Through investigating the robustness of four typical node-similarity-based algorithms in eight real networks, we find that the robustness of link-prediction algorithms for real networks is related to their average degree, while for a network with small world properties, the LP index has the best toughness against the noise among these algorithms. In the end, we prove a method to enhance the robustness of the link-prediction algorithm in network by link weighting, and the result shows it is effective in some for a few networks. At the same time it can improve the precision of algorithms in some biological networks by this way.

## 2. METHOD
### 2.1. Metric

Let $G\ (V, E)$ be an un-weighted undirected network, where $V$ is a set of nodes and $E$ is a set of un-weighted links. The multiple self-connections are excluded from $E$. Every algorithm referred to in this paper assigns a similarity matrix $S$ whose real entry $s_{xy}$ is their similarity score. This score can be viewed as a measure of similarity between nodes $x$ and $y$. For each pair of nodes, $x, y \in V$, generally $s_{xy}=s_{yx}$. All the nonexistent links are sorted in decreasing order according to their scores, and the links at the top are most likely to exist. To test the algorithm's accuracy, a fraction of the observed links $E$ (in this paper always 90% of the whole) is randomly singled out as a training set, $ET$, the remaining links (10% of the whole) are used as the probe set, $EP$, to be predicted and no information in this set is allowed to be used for prediction. Clearly, $E=ET \cup EP$ and $ET \cap EP=\emptyset$. The prediction quality was evaluated by a standard metric, the area under the receiver operating characteristic curve (AUC) [23]. In the present case, this metric can be interpreted as the probability that randomly chosen missing link (a link in $EP$) is given a higher score than a randomly chosen nonexistent link (a link in $U$ but not in $E$, where $U$ denotes the universal set). In the implementation, among $n$ independent comparisons, if there are $n'$ occurrences of missing links having a higher score and $n''$ occurrences of missing links and nonexistent link having the same score, we define the accuracy as:

$$AUC = \frac{n' + 0.5n''}{n} \qquad (1)$$

If all the scores are generated from an independent and identical distribution, the accuracy should be about 0.5. Therefore, the degree to which the accuracy exceeds 0.5 indicates how much better the algorithm performs than pure chance.

In this paper, we calculate the AUC for a network which is treated as without any false links. Then we randomly replace part of links as the false links and then calculate the AUC for this network, which contains a certain proportion of false links, the proportion changed from 0 to 1. We denote the rate of AUC change as $r$, and use it to describe the robustness of link-prediction algorithm, which is denoted as:

$$r = \frac{AUC(f)}{AUC(0)} \qquad (2)$$

where $f$ is the proportion of the false links in a real network. AUC(0) is the AUC for a network without false links.

### 2.2. Similarity indices

We compare the robustness of four similarity indices, including: Common Neighbor (CN), Adamic-Adar (AA) index, Resource Allocation (RA) index and Local Path index (LP) index. Their definitions are as following.

(i) CN. In common sense, two nodes, $x$ and $y$, the more neighbors they have, the more likely to form a link. Let $\Gamma(x)$ denote the set of neighbors of node $x$. The similarity measure of this neighborhood overlap is the directed count:

$$s_{xy} = \left| \Gamma(x) \cap \Gamma(y) \right| \qquad (3)$$

(ii) AA. It refines the simple counting of common neighbors by assigning the less-connected neighbored more weight, as:

$$s_{xy} = \sum_{Z \in \Gamma(x) \cap \Gamma(y)} \frac{1}{\log k(Z)} \tag{4}$$

where $z$ is the common neighbor between nodes $x$ and $y$, $k(z)$ is the degree of node $z$.

(iii) RA. The node $x$ can send resource to $y$, while they are not directed connected. Their common neighbors play the role of transmitters. We assume that each transmitter has a unit of resource and will equally distribute it between all its neighbors. The amount of resource y received is defined as the similarity between $x$ and $y$, which is:

$$s_{xy} = \sum_{Z \in \Gamma(x) \cap \Gamma(y)} \frac{1}{k(Z)} \tag{5}$$

(iv) LP. This index takes consideration of local paths, with wider horizon than CN. It defined as:

$$s_{xy} = A^2 + \varepsilon A^3 \tag{6}$$

where $S$ denotes the similarity matrix, $A$ is the adjacency matrix, $A^2$ denotes the number of the different paths of length 2 connecting nodes $x$ and $y$. Its value is the number of common neighbor between the two nodes. Similarly, $A^3$ is the path counts of length 3 and $\varepsilon$ is a free parameter.

## 3. RESULT AND DISCUSSION

In this paper, eight real networks are drawn from different fields, which include biological networks, social networks and so on. Table 1 summarizes the basic topological feature of those networks. The datasets information of these networks is simply described as follow: (1) C. Elegans' Neural network (Neural) [24, 25]: A directed, weighted network representing the neural network of C. Elegans, which is compiled by D. Watts and S.Strogatz. (2) Food web network (FW) [26]: each species is represented as a node of the network, and a link is placed between two species whenever one of them feeds on the other. The FW considered here is selected from Network Analysis of Trophic Dynamics in South Florida Ecosystems (Florida Bay, Dry Season). (3) Protein-protein interaction network (PPI) [25]: a protein-protein interaction network in budding yeast, each node represents protein and links corresponding to the interactions among proteins. (4) Political Blogs Network (Pblogs) [27]: A directed network of hyperlinks between weblogs on US politics. (5) Internet[2]: A symmetrized snapshot of the structure of the Internet at the level of autonomous systems. (6) Email network (Email) [28]: Enron email network, in which there are two reversed links between each node pair. It indicates that email exchanged and the network is undirected in fact. (7) Power grid network (Grid) [24]: an undirected, un-weighted network representing the topology of the Western States Power Grid of the United States. The data was also compiled by D. Watts and S.Strogatz. (8) Geom Collaboration network in computational geometry (CCG) [29, 30]: an undirected network, which is authors collaboration network based on the file geombib. Two authors are linked with an edge, if they wrote a common work (paper, book ...). The value of an edge is the number of common works.

---

[2] This snapshot was created by Newman M E J from data for July 22, 2006 and is not previously published, available at http://www-personal.unmich.edu/~mejn/netdatd

Table **1**. The statistical properties of six example networks. *N* and *M* are the total numbers of nodes and links. *<k>* is the average degree. *e* is the network efficiency [31]. *C* and *R* are the clustering coefficient [26] and assortative coefficient [32], respectively. *H* is the degree heterogeneity, defined as $<k^2>/<k>^2$.

| Networks | N | M | <k> | e | C | R | H |
|---|---|---|---|---|---|---|---|
| Neural | 297 | 2148 | 14.465 | 0.445 | 0.308 | -0.163 | 1.801 |
| FW | 128 | 2106 | 32.900 | 0.399 | 0.335 | -0.104 | 1.231 |
| PPI | 2361 | 7182 | 5.943 | 0.218 | 0.291 | 0.059 | 2.763 |
| Pblogs | 1224 | 19090 | 27.360 | 0.397 | 0.361 | -0.079 | 3.130 |
| Internet | 22963 | 48436 | 4.219 | 0.276 | 0.350 | -0.199 | 61.978 |
| Email | 36692 | 183831 | 10.020 | 0.221 | 0.716 | -0.111 | 13.980 |
| Grid | 4941 | 6594 | 2.669 | 0.056 | 0.107 | 0.003 | 1.450 |
| CCG | 3621 | 9461 | 5.227 | 0.506 | 0.408 | 0.243 | 4.706 |

In this section, we compare the robustness of four link-prediction algorithms in eight real networks. As shown in Fig. 1, the LP index has the best robustness in most networks. But in Grid and PPI network, the LP index lost its advantage. Especially in Grid network, it even is the worst of all. Comparing their properties, we can find that both the Grid and PPI network, they have lower network efficiency and clustering coefficient than the other networks. Maybe we can say that for a network with high efficiency and clustering coefficient, the LP index has better robustness than the other algorithms based on local information similarity. We have known that LP index performs remarkably better than CN index for making the similarities much more distinguishable by taking account of the contribution of the next-nearest neighbors. Similarly, it also brings the LP index good robustness. When some links are replaced, the similarity score between two nodes of CN index may become 0 for losing their common neighbor, while the LP index may doesn't because of considering the third order neighbors. There are a lot of short loops in networks with high efficiency and clustering coefficient, therefore we can easily find some nodes connecting with short path (such as of length 2 or 3). In practice, most of real complex networks, such as biological network, computer internet network and social network are all of such properties. According to the above discussion, LP index can provide more reliable forecast results in these networks than CN index.

On the other hand, AA index and RA index almost have the same performance as CN index. The reason for this result may be that all of them only consider their common neighbors. Although AA and RA index have better accuracy than CN index in some networks, their robustness are all worse than LP index. In addition, even the LP index can't keep its good robustness along with the increasing of noise strength. When the noise strength exceeds 40% (even more low), the *r* decreases almost linearly. Consequently, if a network has too much false information, the link-prediction algorithms will be invalid.

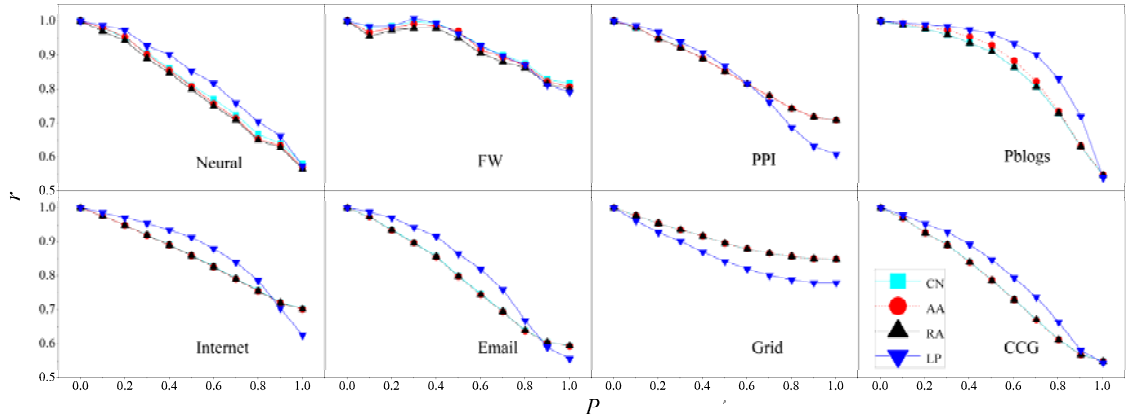

Fig. 1: (Color online) Prediction accuracy vs the strength of randomness for four similarity indices in real networks. There are four kinds of color lines represent four link-prediction algorithms, the CN, AA, RA and LP index. The X-axis $P$ denotes the rate of the false links, and the Y-axis $r$ denotes the rate of the AUC change. Each data point is obtained by average ten independent realizations.

Then we compare the robustness of four kinds of index in different networks respective as shown in Fig. 2. We find that all of the four link-prediction algorithms have good performance in some biological networks, such as FW network, PPI network and Neural network. At the same time, a network with high average degree and degree heterogeneity, such as Pblogs network and Internet network also have good robustness. Both CN index and LP index have a close relationship with the degree, and even all the link-prediction algorithms based on local similarity are due to the degree. Therefore the average degree becomes an import influencing factor. If a network has large average degree, it can cancel out the influence from the error links well. Similarity, if a network has high degree heterogeneity, there are a few nodes with high degree. It can effectively keep the robustness of network structure. This is especially obvious in biological networks. Therefore it can keep the effectiveness of these topological informations to some extent. From this, these algorithms based on local similarity will get convincing results in these networks with larger average degree and degree heterogeneity.

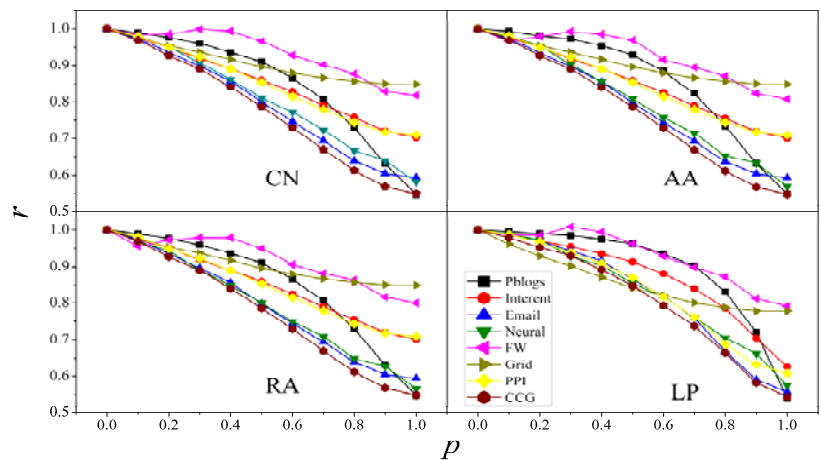

Fig. 2: (Color online) Prediction accuracy vs the strength of randomness of the four algorithms in the eight

different real networks. *P* is the rate of false links in a real network, and *r* denotes the rate of the AUC change. Eight kinds of color lines represent eight different networks respectively. Each data point is obtained by average ten independent realizations.

**4. ENHANCE THE ROBUSTNESS**

In PPI networks, it can effectively eliminate noisy data through weighting network structure. Inspired by this, we make an attempt to enhance the robustness of the link-prediction algorithm by link weighting. First, we give every link a weight, which denotes the probability of existence of the link in real networks. In this paper, we consider the link clustering coefficient as the weight of the links (the structure weights), which is defined as:

$$w_{ij} = \frac{N_{ij}}{\max[k_i, k_j]} \quad (7)$$

where $N_{ij}$ means the number of common neighbors. Here we do not use the real weight of networks for some reasons, the one is that the calculation of the real weights is more complicated than the structure weights, and because of the existing of the Weak-Ties effect [33-35] in link prediction, it becomes more difficult to ensure the most appropriate weights; on the other hand, the structure weights can reflect the topological structure of networks to some extent, these information may be useful for these link-prediction algorithms based on the topological similarity.

Zhou *et al*. [35] extend some similarity index (CN, AA and RA index) from binary networks to weighted networks according to Murata and Moriyasu's reach [36]. And Bai *et al.* first developed the weighted LP index [37]. The four weighted similarity indices are summarized as below.

(i) Weighted CN

$$s_{xy} = \sum_{z \in \Gamma(x) \cap \Gamma(y)} w(x,z) + w(z,y) \quad (8)$$

where $w(x,z)=w(z,x)$ denotes the link weight between nodes *x* and *z*.

(ii) Weighted AA

$$s_{xy} = \sum_{z \in \Gamma(x) \cap \Gamma(y)} \frac{w(x,z) + w(z,y)}{\log(1+s(z))} \quad (9)$$

where $s(x)=\sum_{z \in \Gamma(x)} w(x,z)$ denotes the strength of node *z*.

(iii) Weighted RA

$$S_{xy} = \sum_{z \in \Gamma(x) \cap \Gamma(y)} \frac{w(x,z) + w(z,y)}{s(z)} \quad (10)$$

(iv) Weighted LP

$$s_{xy} = \sum_{z \in \Gamma(x) \cap \Gamma(y)} \left(w(x,z) + w(z,y)\right) + \varepsilon \sum_{i,j \in l_{x \to y}} \left(w(x,i) + w(i,j)\right)\left(w(i,j) + w(j,y)\right) \quad (11)$$

By this way, the robustness of part of algorithm can be enhanced in a few networks, while this improvement is not obvious, such as Neural and FW network, as shown in Fig. 4. Here we don't compare the AA and RA index for they almost have the same robustness with CN index in most networks. At the same time, we find that the precision of the algorithm is improved as shown in table 2, especially for some biological networks, such as the FW and Neural network. And this advantage can be kept with the increasing of noise strength in the case of that the strength is no more than 0.4. In order to explain this result, we apply a motif analysis [38]. There are six different types of four-node connected subgraph in these networks as shown in Fig. 3. The number of motifs-3 in Neural and FW network is more than other networks. Taking the CN index for example, the links 2-3, 2-4 and 3-4 are all signed score one by un-weighted CN index, while the score of links 2-3 and 3-4 become 0.833, link 2-4 is 0.667 by weighted CN index. Thus, the weighted CN index is more distinguishable than the un-weighted one. On the other hand, because of the score of link 2-3 or 3-4 is bigger than 2-4, it is more likely to exist link 1-4 or 1-2. In another word, the motifs-4 and motifs-6 will be more likely to exist than motifs-1, 2 and 5. Calculation results show good agreement with our analysis in these biological networks. Thus we can easily find that in FW network, the weighted CN index even has better performance than un-weighed LP index in Fig. 4. And we find that some biological networks especially Food Webs networks contains a very large number of motifs-3, therefore it may have better results in biological networks by this weighting pattern.

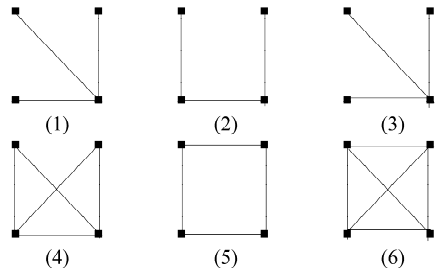

Fig. 3 Six Different types of four-node connected subgraph. We denote the four nodes as 1, 2, 3 and 4 in a clockwise and the node on the top left corner is node 1.

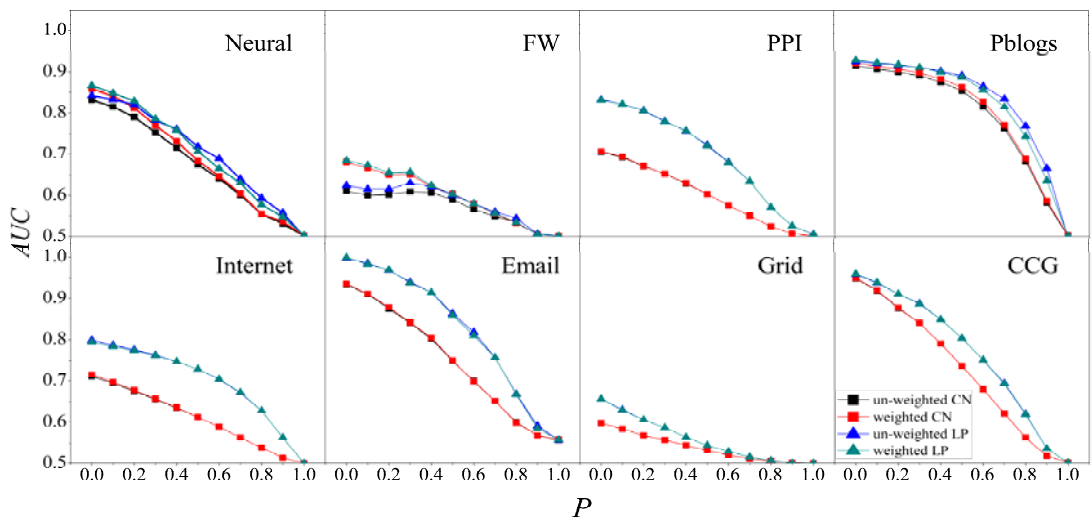

Fig. 4: (Color online) the robustness of the CN and LP index for the real weighted networks and un-weighted ones. The X-axis *P* denotes the rate of the false links, and the Y-axis denotes the AUC change with the increase of the noise strength. Each data point is obtained by average ten independent realizations.

Table **2**: Comparing of the weighted accuracies of algorithms and un-weighted accuracies of algorithms, measured by AUC and *P* is 0. Each number is obtained by averaging over 10 implementations with independently random partitions of testing set and probe set. The parameter for LP, $\varepsilon$, is fixed as $10^{-3}$.

| Network | | | CN | AA | RA | LP |
|---|---|---|---|---|---|---|
| Biological Networks | Neural | un-weighted | 0.850 | 0.871 | 0.875 | 0.865 |
| | | weighted | **0.877** | **0.883** | **0.885** | **0.882** |
| | FW | un-weighted | 0.610 | 0.616 | 0.623 | 0.624 |
| | | weighted | **0.680** | **0.676** | **0.671** | **0.684** |
| | PPI | un-weighted | **0.706** | **0.706** | **0.707** | 0.832 |
| | | weight | **0.707** | **0.707** | 0.707 | 0.831 |
| Other Networks | Pblogs | un-weighted | 0.918 | 0.920 | 0.922 | 0.929 |
| | | weighted | **0.924** | **0.925** | **0.925** | **0.932** |
| | Internet | un-weighted | 0.711 | 0.713 | 0.713 | 0.799 |
| | | weighted | **0.715** | **0.715** | **0.715** | 0.795 |
| | Email | un-weighted | 0.907 | 0.908 | 0.908 | 0.915 |
| | | weighted | **0.908** | **0.909** | **0.908** | **0.917** |
| | Grid | un-weighted | 0.589 | 0.589 | 0.589 | 0.642 |
| | | weighted | 0.589 | 0.589 | 0.589 | 0.642 |
| | CCG | un-weighted | 0.908 | 0.910 | 0.910 | 0.917 |
| | | weight | **0.909** | 0.910 | 0.910 | **0.918** |

## 5. CONCLUSION

In this paper, we study the robustness of four link-prediction algorithms based on local information similarity in eight real networks. In network with small world properties, the LP index has the best robustness among four index based on similarity. But if both network efficiency and clustering coefficient are small, the LP index lost its advantage, *e.g.* the Grid network. The CN, AA and RA index almost have the same robustness. In addition, for different networks, the link-prediction algorithm used for a network with large average degree will have strong robustness. Another influencing factor is the heterogeneity of degree, the larger of the degree heterogeneity a network has, and the better robustness of link-prediction algorithm performs in this network. At the end, we make an attempt to enhance the robustness by link weighting, and find that it is useful for some biological networks, especially Food Webs networks. An additional result is the accuracy of the prediction can be improved by this way.

Although studied some networks in this letter, we need more computation to prove these results and we just give the most widely cite for these results. On the other hand, it is not obvious to improve the robustness by link weighting with the structure weights but may provide a way that we can enhance the robustness by giving a suitable weight to a network. For example, we can iterate the weight mentioned in the text (the link clustering coefficient) until get a stable result, and we see this result as the weight as the links. We can do more work on enhancing the robustness of the link-prediction algorithm.

**CONFLICT OF INTEREST**

None declared

**ACKNOWLEDGEMENT**

This work has been supported by the National Natural Science Foundation of China (Grant Nos. 11147121, 11264012), the Scientific Research Fund of Education Department of Hunan Province of China (Grant No. 11B128), and partly by the Doctor Startup Project of Xiangtan University (Grant No. 10QDZ20).